\documentclass{article}

\usepackage{arxiv}

\usepackage[utf8]{inputenc} 
\usepackage[T1]{fontenc}    
\usepackage{hyperref}       
\usepackage{url}            
\usepackage{booktabs}       
\usepackage{amsfonts}       
\usepackage{nicefrac}       
\usepackage{microtype}      
\usepackage{lipsum}		
\usepackage{graphicx}
\usepackage{amsmath}
\usepackage{amssymb}
\usepackage{color}

\title{Static conformal elastic solution of Einstein's field equations}

\author{H. M. Manjunatha$^{*}$ and S. K. Narasimhamurthy$^{**}$ \\
	Department of PG Studies and Research in Mathematics\\
	Kuvempu University, Jnana Sahyadri\\
	Shankaraghatta - 577451 \\
    Shivamogga, Karnataka, India\\
	$^{*}$\texttt{manjunathahmnmt@gmail.com} \\
    $^{**}$\texttt{nmurthysk@gmail.com} \\
	\And
	Z. Nekouee \\
	Department of Mathematics \\
    Faculty of Mathematical Sciences \\
    University of Mazandaran \\
    P. O. Box 47416-95447, Babolsar, Iran \\
	\texttt{zohrehnekouee@gmail.com} \\
}

\date{}

\renewcommand{\headeright}{}
\renewcommand{\undertitle}{}

\hypersetup{
pdftitle={Static conformal elastic solution of Einstein's field equations},
pdfsubject={gr-qc},
pdfauthor={H. M. Manjunatha, S. K. Narasimhamurthy, Z. Nekouee},
pdfkeywords={Relativistic elasticity, Conformal transformations, Space-time matching},
}

\begin{document}
\maketitle

\begin{abstract}
In this paper, we study new exact solutions of Einstein's field equations with the motivation of the relativistic elasticity theory. We construct the static conformal elastic solution by applying conformal transformations to the Reissner-Nordstr$\ddot{\text{o}}$m-de Sitter solution. The conformal factors for static space-time structure with an elastic matter are obtained. Further, we analyze the viability of energy conditions. And also discuss the matching problem with the exterior space-time.  
\end{abstract}



\section{Introduction}\label{s:1}
\par Nowadays, a lot of research work is going on the properties of neutron stars. Many physicists have discussed the mechanical behavior of solid crusts of neutron stars. Thus, the relativistic theory of elasticity \cite{Carter} has become a popular area of research today. In \cite{Beig}, one can see the formulation of relativistic elasticity as a field theory obtained from a Lagrangian. The Hookean elasticity theory is not appropriate for the explanation of solid matter at high pressures. To resolve this issue, Carter and Quintana \cite{Carter2} described the high-pressure quasi-Hookean approximation, which helps to analyze the interiors of neutron stars.

\par The relativistic theory of elasticity has a lot of applications in astrophysics, for example, in the analysis of deformations of heavier stellar objects. The interplay between the gravitational field and solid elastic material is a significant problem in astrophysical applications. This interplay may regard as the theory of two symmetric tensor fields, such as the physical metric $g$ and the material metric $h$. In 1911, Herglotz studied the relativistic theory of elasticity for special relativity, and in 1916, Nordstr$\ddot{\text{o}}$m discussed it for general relativity \cite{Beig}. Ref. \cite{Kijowski} introduces a new approach to the relativistic elasticity theory. Wherein the diffeomorphisms of the material space play the aspect of gauge transformations.

\par Schwarzschild introduced the first two exact solutions of Einstein's field equations (EFEs). The exterior Schwarzschild solution explains the space-time geometry outside the spherically symmetric mass. And the interior Schwarzschild solution is equivalent to the inside geometry of a fluid sphere that has homogeneous energy density. Tolman introduced five various types of the exact solutions of EFEs with static configuration. One can see the detailed discussion on the exact solutions in \cite{Negi}. Geometric properties of the non-static cylindrical vacuum solution of EFEs are analyzed in \cite{Manjunatha1}. The Finslerian generalized EFEs and their exact solutions are analyzed in \cite{Li} and references therein. The Finslerian wormhole solutions are investigated in \cite{Rahaman-1,Manjunatha2}.

\par Stability theory forms a central platform in the mathematical concepts of general relativity. Many researchers have put a lot of effort into analyzing the stability of exact solutions of the EFEs. Understanding the stability of black holes (BHs) concerning perturbations has more significance in investigating the existence of BHs and their evolution. One can characterize the stability by considering a small perturbation in the metric and computing the variation of EFE. And discuss different types of gravitational stabilities like mode stability, linear stability, and non-linear stability. For most of the exact solutions, non-linear stability is still under discussion. Non-linear stability is investigated in \cite{Hintz} for the Kerr-Newman-de Sitter solution considering small angular momentum. Since BHs regarded as thermodynamic systems, one can also discuss thermodynamic stability with gravitational stabilities \cite{York}. Regge and Wheeler \cite{Regge} found that the Schwarzschild solution was stable concerning weak modal perturbations. Schwarzschild BHs are also found to be stable concerning linear scalar perturbations. Ref. \cite{Dafermos} discusses the general linear stability of the Schwarzschild solution. From \cite{Klainerman}, we found that the Schwarzschild solution is stable concerning axial, polarized, symmetric, non-linear perturbations. The linearization stability of space-times and perturbation theory are studied in \cite{Fischer}. They \cite{Fischer} investigated that space-time is linearization stable if it has no Killing fields. Stability is discussed through linearization of EFE considering Robertson-Walker cosmological model in \cite{Girbau}.

\par Recently, by evaluating the gravitational binding energy, the energetic stability of the exact solutions of EFEs is investigated in \cite{deLyra}. They \cite{deLyra} found that the interior Schwarzschild solution has vanishing binding energy. And hence it is maximally unstable. The $D$-dimensional Schwarzschild BHs and their stability are discussed in \cite{Ishibashi}. The stability of Schwarzschild-de Sitter (SdS) BHs is studied in \cite{Konoplya3} and that of Kaluza-Klein BHs in \cite{Ishihara}. From \cite{Konoplya2}, we found that asymptotically anti-de Sitter (AdS) BHs are stable concerning gravitational perturbations. And also that they are thermodynamically stable for some values of parameters. However, small AdS BHs are thermodynamically unstable within the context of ordinary Einstein-Maxwell theory. Hence they may undergo Hawking-Page transition due to gravitational instability.

\par Conformal transformations are used to construct the new exact solutions of EFEs. In \cite{Loranger}, an explicit four-fold infinity of new physically viable exact perfect fluid solutions of EFEs is generated using conformal transformations. These constructed solutions have a fascinating application in studying the internal characteristics of neutron stars. Brito \textit{et al.} \cite{Brito4} have studied general elastic stars in relativity with spherical symmetry. The conformal transformation between the perfect fluid metric and the exterior Schwarzschild space-time metric is discussed in \cite{Hansraj}.

\par Reissner-Nordstr$\ddot{\text{o}}$m-de Sitter (RNdS) space-time \cite{Stuchlik} is equivalent to the gravitational field generated by the spherically symmetric charged gravitational source. It is a solution of the Einstein-Maxwell equations with a non-vanishing cosmological constant. The Ref. \cite{Carter3} generalizes the BH solution of Reissner-Nordstr$\ddot{\text{o}}$m (RN) type to include the cosmological constant. Reissner-Nordstr$\ddot{\text{o}}$m-(anti-)de Sitter space-times are classified into eleven types \cite{Stuchlik} based on the qualitatively distinct nature of a geodetical motion. The geodetical structure of space-time describes the motion of test particles having no charge and the motion of the photons. It should reduce to the RN solution \cite{Bicak}, and SdS solution \cite{Manjunatha} if $\Lambda=0$, and $Q=0$, respectively. The Cauchy horizon, event horizon, and cosmological horizon of RNdS space-time are discussed in \cite{Chambers,Hollands}. The inner horizon and the event horizon of RN solution are, respectively, future inner trapping horizon and future outer trapping horizon \cite{Faraoni,Jakobsson}.

\par RN solution is stable under linear perturbations considering four dimensions \cite{Moncrief}. Higher-dimensional BHs and their stability are investigated in \cite{Ishibashi2}. From \cite{Konoplya}, we found that RNdS BH becomes unstable at a sufficient enormous charge considering $D>6$ dimensions. Through perturbation analysis, the instability forms and their corresponding thresholds of RNdS BH are analyzed in \cite{Tanabe}. Further, they \cite{Tanabe} investigated that the gravitational perturbation of the RNdS BH becomes unstable. From the expressions of geodesic deviation, the stability of both Schwarzschild solution and RN solution is analyzed in \cite{Nashed}. Recently, we have discussed the thermodynamic features of RNdS (RN) BHs. And also their thermodynamic stability \cite{Manjunatha3}.

\par The features of RNdS space-time will pave the path for us to realize the space-time evolution. And yield a theoretical foundation for discovering the physical mechanism of the universe's accelerated expansion. The evolving process of the RNdS space-time to a pure de Sitter (dS) space-time is different because of the disparate electric potential at the BH horizon. Hence, the existence of electric potential at the BH horizon has an essential role in space-time evolution. The obtained conclusions of the theoretical investigation on the RNdS space-time evolution can be used for simulating the universe evolution, which opens a new path to investigate the internal facts that support the accelerated expansion of the universe \cite{Li2}. Because of this fascinating feature of the RNdS space-time, in this work, we would like to construct the elastic solution of the EFEs by applying conformal transformations to the RNdS solution. Here, we focus on the static configuration. That is, the conformal factor depends on radial coordinate $r$ only.

\par The present paper is composed in the following form. Sec. \ref{s:2} deals with some basic ideas of the relativistic elasticity theory. In Sec. \ref{s:3}, the static conformal elastic solution of the EFEs is constructed from the RNdS solution. In Sec. \ref{ss:3.1}, we study the elastic fluid having non-zero tangential pressure and zero radial pressure in four cases. And in Sec. \ref{ss:3.2}, the elastic fluid with non-zero radial pressure and zero tangential pressure is discussed in four cases. In Sec. \ref{ss:3.3} and Sec. \ref{ss:3.4}, we study an elastic fluid with non-zero radial and tangential pressures corresponding to a particular form of the conformal factor. Sec. \ref{s:4} is devoted to investigating the matching problem with the exterior space-time. Conclusions and future perspectives are given in Sec. \ref{s:5}.

\section{Basic formulations of the relativistic theory of elasticity}\label{s:2}
In this section, we reminiscence some basic concepts of the relativistic theory of elasticity. Given
\begin{equation*}
ds^{2} = g_{ij}dx^{i}dx^{j}
\end{equation*}
is the line-element of space-time $(\mathcal{M}, g)$ and
\begin{equation*}
d\bar{s}^{2} = \bar{g}_{ij}dx^{i}dx^{j}
\end{equation*}
is the line-element of its related space-time $(\mathcal{M}, \bar{g})$. Then the space-times $(\mathcal{M}, g)$ and $(\mathcal{M}, \bar{g})$ are said to be conformally related if
\begin{equation*}
\bar{g}_{ij} = e^{2\sigma}g_{ij}\,\,\,\text{and}\,\,\,\bar{g}^{ij} = e^{-2\sigma}g^{ij},
\end{equation*}
where $\sigma=\sigma(x^{l})$ is a non-zero function $\sigma:\mathcal{M}\rightarrow\mathbb{R}$. Here $x^{l}$ denote the coordinates in $\mathcal{M}$. Now we define the conformal transformation as the transformation connecting metric $g$ and $\bar{g}$.
\par Given $\mathcal{M}$ denotes the general relativistic space-time equipped with the metric tensor $g_{ij}\,(i,\,j\,=0,\,1,\,2,\,3)$ whose signature is $(-,\,+,\,+,\,+)$. The points of a three-dimensional manifold $\mathcal{N}$ describe the material particles making up a continuous medium. We consider that  $y^{A}\,(A\,=1,\,2,\,3)$ are the coordinates in the material manifold $\mathcal{N}$ and $\gamma$ is a material metric. The map $\varphi:\mathcal{M} \rightarrow \mathcal{N}$ describes the space-time structure of the material through the coordinates $y^{A}=y^{A}(x^{l})$ and the map is called a configuration map. These maps describe the states of relativistic continuation. The physical laws explain the mechanical properties of the elastic medium. And governed by the system of hyperbolic second-order PDE. Let $\varphi_{*}: T_{p}\mathcal{M} \rightarrow T_{\varphi(p)}\mathcal{N}$ be the differential map. It gives rise to a matrix of order $3\times4$ with the elements
\begin{equation}\label{e:3.1}
y^{A}_{l} = \frac{\partial y^{A}}{\partial x^{l}}.
\end{equation}
The element (\ref{e:3.1}) denotes the relativistic deformation gradient \cite{Brito3}. Let $u^{l}$ be a time-like, future-oriented unit vector field. It describes the velocity field of the matter. Kernel of the element (\ref{e:3.1}) is of dimension one and is spanned by the unit vector field $u^{l}$. The unit vector field $u^{l}$ satisfies the following conditions:
\begin{equation}\label{e:3.2}
u^{l}y^{A}_{l} = 0,\,\,\,\,u^{l}u_{l}=-1,\,\,\,u^{0}>0.
\end{equation}
The strain tensor is given by
\begin{equation}\label{e:3.3}
s_{ij} = \frac{1}{2}\bigg(h_{ij}-k_{ij}\bigg)\,\,=\,\,\frac{1}{2}\bigg(g_{ij}-K_{ij}\bigg),
\end{equation}
where $h_{ij}=g_{ij}+u_{i}u_{j}$, $K_{ij}=-u_{i}u_{j}+k_{ij}$, and $k_{ij}=(\varphi^{*}\gamma)_{ij}=y^{A}_{i}y^{B}_{j}\gamma_{AB}$ is the pulled-back material metric. If $s_{ij}=0$, then we say that the material is locally relaxed. The material metric $\gamma$ determines the distance between particles in a material if it is in the locally relaxed state. Let $p_{1}^{2}, p_{2}^{2}$, and $p_{3}^{2}$ be the eigenvalues of $k^{l}_{i}$. For the elastic deformation, the internal energy $v$ can be expressed as $v=v(\lambda_{1}, \lambda_{2}, \lambda_{3})$, where $\lambda_{1}, \lambda_{2}, \lambda_{3}$ are the invariants of $K$ given by \cite{Brito2}
\begin{equation}\label{e:3.4}
\lambda_{1} = \frac{1}{2}\bigg(\text{tr}\,K-4\bigg),\,\,\,\,\lambda_{2}\,=\,\frac{1}{4}\bigg[\text{tr}\,K^{2}-(\text{tr}\,K)^{2}\bigg]+3,\,\,\,\,\lambda_{3}\,=\,
\frac{1}{2}\bigg(\text{det}\,K-1\bigg).
\end{equation}
Here, we have
\begin{equation}\label{e:3.5}
\text{tr}\,K = 1+p_{1}^{2}+p_{2}^{2}+p_{3}^{2},\,\,\,\text{tr}\,K^{2}\,=\,1+p_{1}^{4}+p_{2}^{4}+p_{3}^{4},\,\,\,\text{det}\,K\,=\,p_{1}^{2}p_{2}^{2}p_{3}^{2}.
\end{equation}
We found that the internal energy $v$ depends on invariants $\lambda_{1}$, $\lambda_{2}$, $\lambda_{3}$ of $K$. Let $\epsilon$ and $\epsilon_{0}$ be the particle number density concerning the volume form corresponding to $h_{ij}$ and $k_{ij}$, respectively. Then, the energy density $\rho$ is given by
\begin{equation}\label{e:3.6}
\rho = \epsilon v(\lambda_{1}, \lambda_{2}, \lambda_{3})\,=\,\epsilon_{0}\sqrt{\text{det}K}v(\lambda_{1}, \lambda_{2}, \lambda_{3}).
\end{equation}
From Eqs. (\ref{e:3.4}) and (\ref{e:3.5}), the internal energy $v$ of elastic deformations can be seen as a function of eigenvalues of $K$, that is, $v=v(p_{1}^{2}, p_{2}^{2}, p_{3}^{2})$.
\par The EFE with the cosmological constant $\Lambda$ can be written in relativistic units $(c=G=1)$ as follows:
\begin{equation}\label{e:3.8}
G_{ij}+\Lambda g_{ij} = -8\pi T_{ij},
\end{equation}
where $T_{ij}$ is the energy-momentum tensor which can be divided into two contributions $T_{ij}{}^{\text{E}}$ and $T_{ij}{}^{\text{EM}}$ \cite{Islam,Medina,Lemos},
\begin{equation}\label{e:3.8(a)}
T_{ij} = T_{ij}{}^{\text{E}}+T_{ij}{}^{\text{EM}}.
\end{equation}
The first part $T_{ij}{}^{\text{E}}$ is material energy-momentum tensor and is considered as an elastic matter for the purpose of the present work. The tensor $T^{l}_{i}{}^{\text{E}}$ is evaluated as follows: \cite{Brito}
\begin{equation}\label{e:3.7}
T^{l}_{i}{}^{E} = -\rho\delta^{l}_{i}+\frac{\partial\rho}{\partial \lambda_{3}}\text{det}K\,h^{l}_{i}-\left(\text{Tr}\,K\,\frac{\partial\rho}{\partial \lambda_{2}}-\frac{\partial\rho}{\partial \lambda_{1}}\right)k^{l}_{i}+\frac{\partial\rho}{\partial\lambda_{2}}k^{l}_{m}k^{m}_{i}.
\end{equation}
The second part $T_{ij}{}^{\text{EM}}$ denotes electromagnetic energy-momentum tensor. The expression for $T_{ij}{}^{\text{EM}}$ is given by \cite{Islam}:
\begin{equation}\label{e:3.8(b)}
T_{ij}{}^{\text{EM}} = -\frac{1}{4\pi}\left(F^{m}_{i}F_{jm}-\frac{1}{4}g_{ij}F_{ml}F^{ml}\right),
\end{equation}
where $F_{ij}$ is the electromagnetic field tensor (Faraday-Maxwell tensor) and can be written as follows:
\begin{equation}\label{e:3.8(c)}
F_{ij} = \nabla_{i}\mathcal{A}_{j}-\nabla_{j}\mathcal{A}_{i},
\end{equation}
where $\mathcal{A}_{i}$ is the electromagnetic gauge potential, and $\nabla_{i}$ denotes covariant derivative. Faraday-Maxwell tensor $F_{ij}$ is related to the current density $J^{i}$, given by Maxwell equations,
\begin{equation}\label{e:3.8(d)}
\nabla_{j}F^{ij} = -4\pi J^{i}.
\end{equation}
Let $\phi(r)$ denotes the electric potential. Then, the electromagnetic gauge potential for the static and spherically symmetric space-time is given by
\begin{equation}\label{e:3.8(e)}
\mathcal{A}_{i} = -\phi(r)\delta^{t}_{i}.
\end{equation}
The electromagnetic field tensor $F_{ij}$ [Eq. (\ref{e:3.8(c)})] satisfies the antisymmetric property $F_{ij}=-F_{ji}$ and hence $F_{ii}=0$. Further, $F_{ij}$ has six independent components. By considering Eq. (\ref{e:3.8(e)}), we obtain only one non-zero independent component given by
\begin{equation}\label{e:3.8(f)}
F_{tr}=\phi'(r).
\end{equation}
Here and throughout this article, prime indicates the radial differential coefficients. From antisymmetric property, we have $F_{rt}=-\phi'(r)$.

\section{Static conformal elastic Reissner-Nordstr$\ddot{\text{o}}$m-de Sitter solution}\label{s:3}
The metric for RNdS solution in Schwarzschild coordinates $(t,r,\theta,\phi)$ with relativistic units $(c=G=1)$ is as follows:
\begin{equation}\label{e:3.9}
ds^{2} = -\xi(r)dt^{2}+\xi(r)^{-1}dr^{2}+r^{2}d\Omega^{2},
\end{equation}
where $\xi(r)=1-\frac{2M}{r}-\frac{1}{3}\Lambda r^{2}+\frac{Q^{2}}{r^{2}}$ and $d\Omega^{2}=d\theta^{2}+\sin^{2}\theta d\phi^{2}$. Here, $\Lambda$ denotes cosmological constant, $M$ denotes spherically symmetric mass, and $Q$ denotes electric charge.
\par Let $g_{ij}$ and $\bar{g}_{ij}$ denote the elastic fluid metric with spherical symmetry and RNdS space-time metric, respectively. Now we consider that the metrics $g_{ij}$ and $\bar{g}_{ij}$ are conformally related. So we have
\begin{equation}\label{e:3.10}
g_{ij} = f^{2}(r)\bar{g}_{ij},
\end{equation}
with the line-element
\begin{equation}\label{e:3.11}
ds^{2} = f^{2}(r)\left[-\xi(r)dt^{2}+\xi(r)^{-1}dr^{2}+r^{2}d\Omega^{2}\right].
\end{equation}
In Eq. (\ref{e:3.10}), $f=f(r)$ is a static conformal factor, which is to be determined. The Eq. (\ref{e:3.11}) is the static conformal elastic RNdS solution of EFEs.
\par The space-time coordinates are $x^{0}=t, x^{1}=r, x^{2}=\theta, x^{3}=\phi$, and the material coordinates in the material manifold $\mathcal{N}$ are $y^{1}=\tilde{r}, y^{2}=\tilde{\theta}, y^{3}=\tilde{\phi}$, in the context of relativistic theory of elasticity. Given $d\tilde{\Omega}^{2}=d\tilde{\theta}^{2}+\sin^{2}\tilde{\theta}d\tilde{\phi}^{2}$, now we assume that the material metric $\gamma_{AB}$ in $\mathcal{N}$ is flat, so that, its line-element is given by
\begin{equation}\label{e:3.12}
d\Sigma^{2} = d\tilde{r}^{2}+\tilde{r}^{2}d\tilde{\Omega}^{2}.
\end{equation}
Due to spherical symmetry, we can consider the material coordinates as $(\tilde{r}, \tilde{\theta}, \tilde{\phi}) = (\tilde{r}(r), \theta, \phi)$. Hence, we have $d\tilde{\Omega}^{2}=d\Omega^{2}$. Thus Eq. (\ref{e:3.12}) becomes
\begin{equation}\label{e:3.14}
d\Sigma^{2} = d\tilde{r}^{2}+\tilde{r}^{2}d\Omega^{2}.
\end{equation}
The pulled-back material metric for the RNdS space-time admitting conformal transformation is given by
\begin{align}
\label{e:3.15} k^{l}_{i} &= g^{lm}k_{mi} = g^{lm}\gamma _{LJ}y^{L}_{m}y^{J}_{i} \\
\label{e:3.16} &= f^{-2}\xi(r)\tilde{r}'^{2}\delta^{l}_{1}\delta^{1}_{i}+
r^{-2}f^{-2}\tilde{r}^{2}(\delta^{l}_{2}\delta^{2}_{i}+\delta^{l}_{3}\delta^{3}_{i}).
\end{align}
The eigenvalues $\alpha$ and $\beta$ of the metric $k^{l}_{i}$ are given by
\begin{equation}\label{e:3.17}
\alpha = f^{-2}\xi(r)\tilde{r}'^{2},\,\,\,\,\beta=r^{-2}f^{-2}\tilde{r}^{2}.
\end{equation}
The eigenvalue $\beta$ has the algebraic multiplicity two. One must consider the range of radial coordinate $r$ such that eigenvalues of the metric $k^{l}_{i}$ are positive.
\par Let $u^{l}=\left(f^{-1}\xi(r)^{-\frac{1}{2}}, 0, 0, 0\right)$ be the velocity field of matter. We consider $\alpha=p_{1}^{2}$ and $\beta=p_{2}^{2}=p_{3}^{2}$. Then, using Eqs. (\ref{e:3.4}) and (\ref{e:3.5}), the invariants of $K^{l}_{i}=-u^{l}u_{i}+k^{l}_{i}$ are as follows:
\begin{align}
\label{e:3.18} \lambda_{1} &= \frac{1}{2}(\alpha+2\beta-3), \\
\label{e:3.19} \lambda_{2} &= -\frac{1}{2}(\beta^{2}+\alpha+2\beta+2\alpha\beta)+3, \\
\label{e:3.20} \lambda_{3} &= \frac{1}{2}(\alpha\beta^{2}-1).
\end{align}
The invariants of $K^{l}_{i}$ are obtained in terms of $\alpha$ and $\beta$ [see Eqs. (\ref{e:3.18})-(\ref{e:3.20})]. By using Eq. (\ref{e:3.7}), the elastic energy-momentum tensor components are expressed as follows:
\begin{align}
\label{e:3.21} T^{0}_{0}{}^{\text{E}} &= -\rho, \\
\label{e:3.22} T^{1}_{1}{}^{\text{E}} &= -\rho+\frac{\partial\rho}{\partial \lambda_{3}}\,\alpha\beta^{2}-\frac{\partial\rho}{\partial \lambda_{2}}(1+2\beta)\alpha+\frac{\partial\rho}{\partial \lambda_{1}}\alpha, \\
\label{e:3.23} T^{2}_{2}{}^{\text{E}} &= T^{3}_{3}{}^{\text{E}} = -\rho+\frac{\partial\rho}{\partial \lambda_{3}}\,\alpha\beta^{2}-\frac{\partial\rho}{\partial \lambda_{2}}(1+\alpha+\beta)\beta+\frac{\partial\rho}{\partial \lambda_{1}}\beta.
\end{align}
The energy density is given by
\begin{equation}\label{e:3.24}
\rho = \epsilon v(\lambda_{1}, \lambda_{2}, \lambda_{3}) = \epsilon_{0}\beta\sqrt{\alpha}v(\lambda_{1}, \lambda_{2}, \lambda_{3}).
\end{equation}
Using the expressions (\ref{e:3.18})-(\ref{e:3.20}), we obtain that
\begin{align}
\label{e:3.25} \frac{\partial\rho}{\partial\alpha} &= \frac{1}{2}\,\frac{\partial\rho}{\partial \lambda_{1}}-\frac{1}{2}(1+2\beta)\frac{\partial\rho}{\partial \lambda_{2}}+\frac{1}{2}\beta^{2}\frac{\partial\rho}{\partial\lambda_{3}}, \\
\label{e:3.26} \frac{\partial\rho}{\partial\beta} &= \frac{\partial\rho}{\partial \lambda_{1}}-(\alpha+\beta+1)\frac{\partial\rho}{\partial \lambda_{2}}+\alpha\beta\frac{\partial\rho}{\partial\lambda_{3}}.
\end{align}
By plugging Eqs. (\ref{e:3.24})-(\ref{e:3.26}) into Eqs. (\ref{e:3.21})-(\ref{e:3.23}), the elastic energy-momentum tensor components become
\begin{align}
\label{e:3.27} T^{0}_{0}{}^{\text{E}} &= -\epsilon v, \\
\label{e:3.28} T^{1}_{1}{}^{\text{E}} &= 2\epsilon\alpha\frac{\partial v}{\partial\alpha}, \\
\label{e:3.29} T^{2}_{2}{}^{\text{E}} &= T^{3}_{3}{}^{\text{E}} = \epsilon\beta\frac{\partial v}{\partial\beta}.
\end{align}
The electromagnetic energy-momentum tensor components for the RNdS space-time admitting conformal transformation, with elastic configuration and the electromagnetic gauge potential chosen above [Eq. (\ref{e:3.8(e)})], are given by
\begin{align}
\label{e:3.29(a)} T^{0}_{0}{}^{\text{EM}} &= \frac{Q^{2}}{8\pi r^{4}f^{4}},\,\,\,\,T^{1}_{1}{}^{\text{EM}}\,=\,\frac{Q^{2}}{8\pi r^{4}f^{4}}, \\
\label{e:3.29(b)} T^{2}_{2}{}^{\text{EM}} &= -\frac{Q^{2}}{8\pi r^{4}f^{4}},\,\,\,\,T^{3}_{3}{}^{\text{EM}}\,=\,-\frac{Q^{2}}{8\pi r^{4}f^{4}}.
\end{align}
We can easily verify that $T=\text{tr}\,T^{l}_{i}{}^{\text{EM}}=0$. From Eq. (\ref{e:3.8(a)}), the energy-momentum tensor components can be written as follows:
\begin{align}
\label{e:3.29(c)} T^{0}_{0} &= -\epsilon v+\frac{Q^{2}}{8\pi r^{4}f^{4}},\,\,\,\,T^{1}_{1}\,=\,2\epsilon\alpha\frac{\partial v}{\partial\alpha}+\frac{Q^{2}}{8\pi r^{4}f^{4}}, \\
\label{e:3.29(d)} T^{2}_{2} &= T^{3}_{3}\,=\,\epsilon\beta\frac{\partial v}{\partial\beta}-\frac{Q^{2}}{8\pi r^{4}f^{4}}.
\end{align}
For the static RNdS space-time metric (\ref{e:3.11}) admitting conformal transformation, EFEs with elastic matter distribution are obtained as follows:
\begin{align}
\nonumber &G^{0}_{0}+\Lambda = -8\pi T^{0}_{0}: \\
\nonumber &3\Lambda r^{4}f^{4}+3Q^{2}-\bigg[2r^{6}ff''\Lambda-r^{6}f'^{2}\Lambda+6r^{5}ff'\Lambda+3r^{4}f^{2}\Lambda-6r^{2}ff''Q^{2}+3r^{2}f'^{2}Q^{2}-6rff'Q^{2}+3f^{2}Q^{2} \\
\label{e:3.30} &\,\,+12r^{3}ff''M-6r^{3}f'^{2}M+18r^{2}ff'M-6r^{4}ff''+3r^{4}f'^{2}-12r^{3}ff'\bigg] = 24\pi\epsilon vr^{4}f^{4},
\end{align}
\begin{align}
\nonumber &G^{1}_{1}+\Lambda = -8\pi T^{1}_{1}: \\
\nonumber &\Lambda r^{4}f^{4}+Q^{2}-\bigg[r^{6}f'^{2}\Lambda+2r^{5}ff'\Lambda+r^{4}f^{2}\Lambda-3r^{2}f'^{2}Q^{2}-2rff'Q^{2}+f^{2}Q^{2}+6r^{3}f'^{2}M+6r^{2}ff'M-3r^{4}f'^{2} \\
\label{e:3.31} &\,\,-4r^{3}ff'\bigg] = -16\pi\epsilon\alpha\,\frac{\partial v}{\partial\alpha}\,r^{4}f^{4}, 
\end{align}
\begin{align}
\nonumber &G^{2}_{2}+\Lambda = -8\pi T^{2}_{2}: \\
\nonumber &3\Lambda r^{4}f^{4}-3Q^{2}-\bigg[2r^{6}ff''\Lambda-r^{6}f'^{2}\Lambda+6r^{5}ff'\Lambda+3r^{4}f^{2}\Lambda-6r^{2}ff''Q^{2}+3r^{2}f'^{2}Q^{2}+6rff'Q^{2}-3f^{2}Q^{2} \\
\label{e:3.32} &\,\,+12r^{3}ff''M-6r^{3}f'^{2}M-6r^{4}ff''+3r^{4}f'^{2}-6r^{3}ff'\bigg] = -24\pi\epsilon\beta\,\frac{\partial v}{\partial\beta}\,r^{4}f^{4}.
\end{align}
\par The expressions for the radial pressure $p_{r}$ and the tangential pressure $p_{t}$ are respectively, as follows:
\begin{equation}\label{e:3.33}
p_{r} = 2\epsilon\alpha\frac{\partial v}{\partial\alpha},\,\,\,\,p_{t}=\epsilon\beta\frac{\partial v}{\partial\beta}.
\end{equation}
Now, we set $U=\text{ln}\,\alpha$ and $V=\text{ln}\,\beta$. Divide Eqs. (\ref{e:3.31}) and (\ref{e:3.32}) through by (\ref{e:3.30}) then we get
\begin{align}
\nonumber &\bigg[\Lambda r^{4}\bigg\{2rf(rf''+3f')-r^{2}f'^{2}-3f^{2}(f^{2}-1)\bigg\}+3Q^{2}\bigg\{(f-rf')^{2}-2r^{2}ff''-1\bigg\} \\
\nonumber &\,\,\,\,+3r^{2}\bigg\{2ff'(-2r+3M)+rf'^{2}(r-2M)-2rff''(r-2M)\bigg\}\bigg]\,\frac{\partial\,(\text{ln}\,v)}{\partial U} \\
\nonumber &\,\,\,\,=-\frac{3}{2}\bigg[\Lambda r^{4}\bigg\{(f+rf')^{2}-f^{4}\bigg\}+Q^{2}\bigg\{(f-rf')^{2}-4r^{2}f'^{2}-1\bigg\}+f'r^{2} \\
\label{e:3.34} &\,\,\,\,\bigg\{2f(-2r+3M)-3rf'(r-2M)\bigg\}\bigg], 
\end{align}
\begin{align}
\nonumber &\bigg[\Lambda r^{4}\bigg\{2rf(rf''+3f')-r^{2}f'^{2}-3f^{2}(f^{2}-1)\bigg\}+3Q^{2}\bigg\{(f-rf')^{2}-2r^{2}ff''-1\bigg\} \\
\nonumber &\,\,\,\,+3r^{2}\bigg\{2ff'(-2r+3M)+rf'^{2}(r-2M)-2rff''(r-2M)\bigg\}\bigg]\,\frac{\partial\,(\text{ln}\,v)}{\partial V} \hspace{0.4cm} \\
\nonumber &\,\,\,\,=-\bigg[\Lambda r^{4}\bigg\{2rf(rf''+3f')-r^{2}f'^{2}-3f^{2}(f^{2}-1)\bigg\}+3Q^{2}\bigg\{(r^{2}f'^{2}-f^{2}) \\
\label{e:3.35} &\,\,\,\,-2rf(rf''-f')+1\bigg\}+3r^{3}\bigg\{(r-2M)(f'^{2}-2ff'')-2ff'\bigg\}\bigg]. 
\end{align}
From (\ref{e:3.17}), we obtain
\begin{align}
\label{e:3.36} U &= \text{ln}\,\xi(r)+2\,\text{ln}\,\tilde{r}'-2\,\text{ln}\,f, \\
\label{e:3.37} V &= -2\,\text{ln}\,f+2\,\text{ln}\,\tilde{r}-2\,\text{ln}\,r.
\end{align}
From (\ref{e:3.36}) and (\ref{e:3.37}), also we have,
\begin{equation}\label{e:3.38}
U = V+\,\text{ln}\,(3r^{2}-6rM-\Lambda r^{4}+3Q^{2})-\text{ln}\,3\tilde{r}^{2}+2\,\text{ln}\,\tilde{r}'.
\end{equation}
Let $W=\text{ln}\,v$. Then $W=W(U, V)$, and it is clear that $\frac{\partial W}{\partial U}$ and $\frac{\partial W}{\partial V}$ are functions of $r$. Because of a constitutive equation $v=v(\alpha, \beta)$, correspondingly $W(U, V)=\text{ln}\,v(U, V)$, exists, and the property $$\frac{\partial^{2}W}{\partial U\,\partial V}=\frac{\partial^{2}W}{\partial V\,\partial U},$$ must be satisfied. Therefore we have
\begin{equation}\label{e:3.39}
C\left(\frac{3rM-\Lambda r^{4}-3Q^{2}}{r(3r^{2}-6rM-\Lambda r^{4}+3Q^{2})}-\frac{f'}{f}+\frac{\tilde{r}''}{\tilde{r}'}\right) = -\frac{f'}{f}-\frac{1}{r}+\frac{\tilde{r}'}{\tilde{r}},
\end{equation}
where $C\in\mathbb{R}^{+}$. The solution of the ordinary differential equation (\ref{e:3.39}) is given by
\begin{equation}\label{e:3.40}
f(r) = \bar{a}\,\frac{(3r^{2}-6rM-\Lambda r^{4}+3Q^{2})^{\frac{C}{2(C-1)}}}{r}\,(\tilde{r})^{\frac{1}{1-C}}\,(\tilde{r}')^{\frac{C}{C-1}},
\end{equation}
where $\bar{a}\in\mathbb{R}$ and $C\neq1$. Therefore, the conformal factor is given by
\begin{equation}\label{e:3.41}
f^{2}(r) = \bar{a}^{2}\,\frac{(3r^{2}-6rM-\Lambda r^{4}+3Q^{2})^{\frac{C}{C-1}}}{r^{2}}\,(\tilde{r})^{\frac{2}{1-C}}\,(\tilde{r}')^{\frac{2C}{C-1}}.
\end{equation}
Thus, the static conformal elastic RNdS solution of EFEs is given by (\ref{e:3.11}) with $f^{2}(r)$ given by (\ref{e:3.41}).
\par Next, we will discuss the static conformal elastic particular solutions through the following examples:

\subsection{Example 1}\label{ss:3.1}
Now, we study the elastic fluid with $p_{r}=0$ and $p_{t}\neq0$. Since $p_{r}=0$, from Eq. (\ref{e:3.31}), we get
\begin{align}
\nonumber &r^{6}f'^{2}\Lambda+2r^{5}ff'\Lambda+r^{4}f^{2}\Lambda-3r^{2}f'^{2}Q^{2}-2rff'Q^{2}+f^{2}Q^{2}+6r^{3}f'^{2}M+6r^{2}ff'M-3r^{4}f'^{2}-4r^{3}ff'\\
\label{e:3.54} &\,\,\,\, = \Lambda r^{4}f^{4}+Q^{2}. 
\end{align}
\par The first-order ordinary differential equation (\ref{e:3.54}) is very long. Hence we take up the graphical interpretation. Although it is impossible to determine an explicit expression for $f$ by determining the analytical solution of Eq. (\ref{e:3.54}), it is desirable to find the numerical solution for $f$ by providing the initial condition (IC). The Eq. (\ref{e:3.54}) can be solved numerically with IC: $f(2.1)=1.5$ in the following four different cases. IC is chosen at $r=2.1$ since eigenvalues $\alpha$ and $\beta$ of the metric $k^{l}_{i}$ are positive at this point. And with this IC, the problem cannot be initially singular. Further, it should be noted that the RKF45 method is employed to find the numerical solution.
	
\subsubsection{Case 1: $\Lambda\neq0$ and $Q\neq0$}\label{sss:3.11}
\begin{figure}
	\centering
	\includegraphics[width=3.8 in]{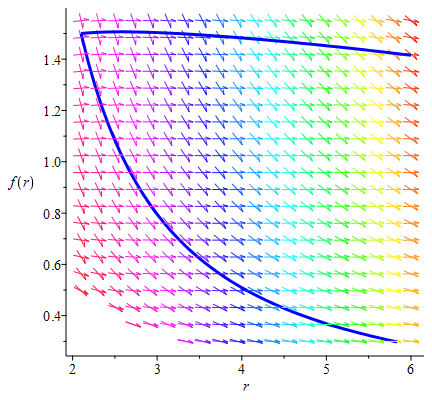}
	\caption{The conformal factor, $f=f(r)$, of the elastic RNdS solution with $p_{r}=0$, $\Lambda=0.01$, $Q=1$, $M=1$ and IC: $f(2.1)=1.5$}
	\label{f:3.1}
\end{figure}

The direction field plot of Eq. (\ref{e:3.54}) is shown in Fig.\,\ref{f:3.1}. Here we consider $\Lambda=0.01, Q=1, M=1$. Fig.\,\ref{f:3.1} describes conformal factor of the static conformal elastic RNdS solution (\ref{e:3.11}) of EFE (\ref{e:3.8}) considering radial pressure $p_{r}=0$.
	
\subsubsection{Case 2: $\Lambda=0$ and $Q\neq0$}\label{sss:3.12}
\begin{figure}
	\centering
	\includegraphics[width=3.8 in]{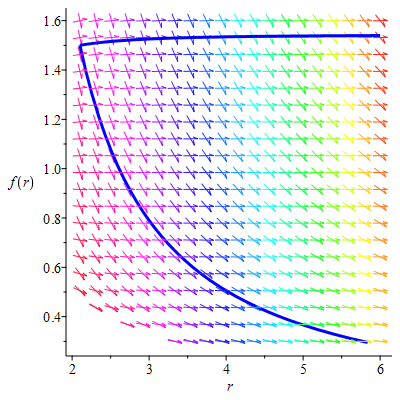}
	\caption{The conformal factor, $f=f(r)$, of the elastic RN solution with $p_{r}=0$, $Q=1$, $M=1$ and IC: $f(2.1)=1.5$}
	\label{f:3.2}
\end{figure}

In this case, Eq. (\ref{e:3.54}) reduces to
\begin{equation}\label{e:3.55}
-3r^{2}f'^{2}Q^{2}-2rff'Q^{2}+f^{2}Q^{2}+6r^{3}f'^{2}M+6r^{2}ff'M-3r^{4}f'^{2}-4r^{3}ff' = Q^{2}.
\end{equation}
Fig.\,\ref{f:3.2} shows the direction field plot of Eq. (\ref{e:3.55}) and describes the conformal factor of the elastic RN solution obtained by applying a conformal transformation to the RN solution considering the elastic matter. In the plot we consider $Q=1$ and $M=1$.

\subsubsection{Case 3: $\Lambda\neq0$ and $Q=0$}\label{sss:3.13}
\begin{figure}
	\centering
	\includegraphics[width=3.8 in]{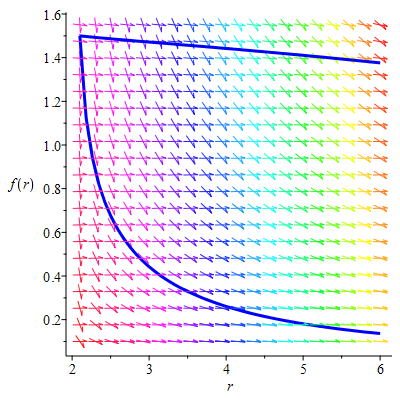}
	\caption{The conformal factor, $f=f(r)$, of the elastic SdS solution with $p_{r}=0$, $\Lambda=0.01$, $M=1$ and IC: $f(2.1)=1.5$}
	\label{f:3.3}
\end{figure}

Now, Eq. (\ref{e:3.54}) reduces to
\begin{equation}\label{e:3.56}
r^{6}f'^{2}\Lambda+2r^{5}ff'\Lambda+r^{4}f^{2}\Lambda+6r^{3}f'^{2}M+6r^{2}ff'M-3r^{4}f'^{2}-4r^{3}ff' = \Lambda r^{4}f^{4}.
\end{equation}
The direction field plot of Eq. (\ref{e:3.56}) is shown in Fig.\,\ref{f:3.3}. Here we consider $M=1$ and $\Lambda=0.01$. Fig.\,\ref{f:3.3} represents the conformal factor of the elastic SdS solution obtained by applying conformal transformation to the SdS solution considering the elastic matter.

\subsubsection{Case 4: $\Lambda=0$ and $Q=0$}\label{sss:3.14}
\begin{figure}
	\centering
	\includegraphics[width=3.8 in]{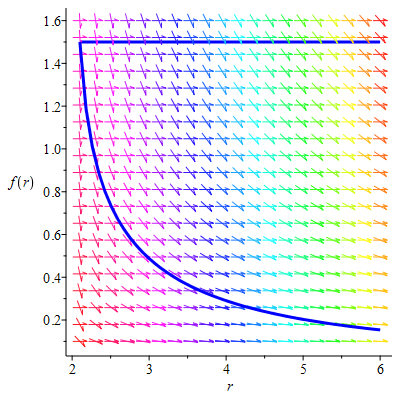}
	\caption{The conformal factor, $f=f(r)$, of the elastic Schwarzschild solution with $p_{r}=0$, $M=1$ and IC: $f(2.1)=1.5$}
	\label{f:3.4}
\end{figure}

Eq. (\ref{e:3.54}) reduces to
\begin{equation}\label{e:3.57}
6r^{3}f'^{2}M+6r^{2}ff'M-3r^{4}f'^{2}-4r^{3}ff' = 0.
\end{equation}
One can see the detailed discussion for this case in \cite{Brito}, where the analytical solution of Eq. (\ref{e:3.57}) is given. Fig.\,\ref{f:3.4} demonstrates the conformal factor of the elastic Schwarzschild solution obtained by applying a conformal transformation to the Schwarzschild solution considering the elastic matter. Hence it describes the numerical solution of the differential equation. Here we consider $M=1$.
\par
\begin{figure}
	\centering
	\includegraphics[width=3.8 in]{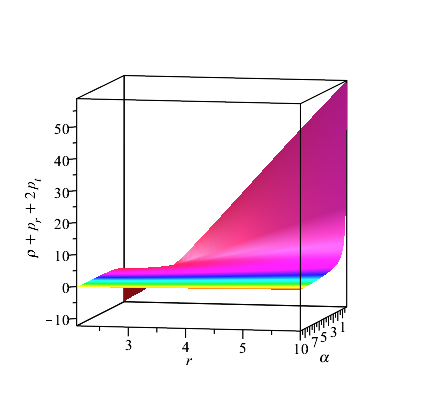}
	\caption{$\rho+p_{r}+2p_{t}=(\rho+p_{r}+2p_{t})(r, \alpha)$ with $M=1$ (The SEC term). Here $\alpha\in\mathbb{R}^{+}$ (see Ref. \cite{Brito})}
	\label{f:3.4(a)}
\end{figure}

 From \cite{Brito}, we have found that the energy density $\rho$ is positive for $r>3M$, and the dominant energy condition (DEC) ($\rho-|p_{t}|\geq0, \rho-|p_{r}|=\rho\geq0$) is satisfied for particular ranges of $r>3M$. Hence it is clear that both weak energy condition (WEC) and null energy condition (NEC) are validated for particular ranges of $r>3M$. The validation of the strong energy condition (SEC) is analyzed through Fig.\,\ref{f:3.4(a)}, wherein we consider $M=1$ and $\alpha\in\mathbb{R}^{+}$ (see Ref. \cite{Brito}). From Fig.\,\ref{f:3.4(a)}, the SEC term $\rho+p_{r}+2p_{t}$ is obtained to be negative for $r<3M$ and positive for $r>3M$. So, we conclude that the SEC is satisfied for $r>3M$.
	
\subsection{Example 2}\label{ss:3.2}
In this section, we have discussed an elastic fluid with $p_{r}\neq0$ and $p_{t}=0$. Since $p_{t}=0$, the following second-order ordinary differential equation is obtained from Eq. (\ref{e:3.32}):
\begin{align}
\nonumber &2r^{6}ff''\Lambda-r^{6}f'^{2}\Lambda+6r^{5}ff'\Lambda+3r^{4}f^{2}\Lambda-6r^{2}ff''Q^{2}+3r^{2}f'^{2}Q^{2}+6rff'Q^{2}-3f^{2}Q^{2}+12r^{3}ff''M \\
\label{e:3.58} &\,\,\,\,-6r^{3}f'^{2}M-6r^{4}ff''+3r^{4}f'^{2}-6r^{3}ff' = 3\Lambda r^{4}f^{4}-3Q^{2}. 
\end{align}
\par The analytical solution of Eq. (\ref{e:3.58}) has a complicated structure. So we have undertaken the analysis by graphical interpretation and studied the numerical solution of Eq. (\ref{e:3.58}). The Eq. (\ref{e:3.58}) can be solved numerically with ICs: $f(2.1)=1.5, f'(2.1)=0.5$ in the following four different cases. As in the previous example, ICs are chosen at $r=2.1$ since eigenvalues $\alpha$ and $\beta$ of the metric $k^{l}_{i}$ are positive at this value of $r$. And with these ICs, the problem cannot be initially singular. Here also, the RKF45 method is employed to find the numerical solution.
	
\subsubsection{Case 1: $\Lambda\neq0$ and $Q\neq0$}\label{sss:3.21}
\begin{figure}
	\centering
	\includegraphics[width=3.8 in]{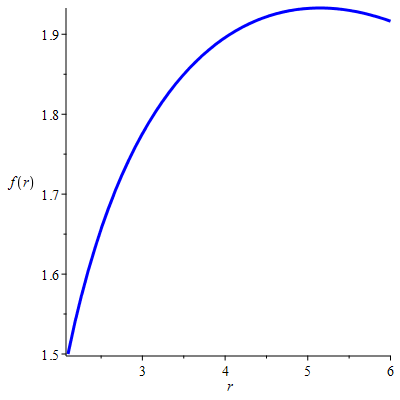}
	\caption{The conformal factor $f(r)$ against $r$ for the elastic RNdS solution considering $p_{t}=0$, $\Lambda=0.01$, $Q=1$, $M=1$ and ICs: $f(2.1)=1.5$, $f'(2.1)=0.5$}
	\label{f:3.5}
\end{figure}

The numerical solution of Eq. (\ref{e:3.58}) is shown in Fig.\,\ref{f:3.5}. Here we consider $\Lambda=0.01, Q=1, M=1$. Fig.\,\ref{f:3.5} describes conformal factor of the static conformal elastic RNdS solution (\ref{e:3.11}) of EFE (\ref{e:3.8}) considering tangential pressure $p_{t}=0$.
	
\subsubsection{Case 2: $\Lambda=0$ and $Q\neq0$}\label{sss:3.22}
\begin{figure}
	\centering
	\includegraphics[width=3.8 in]{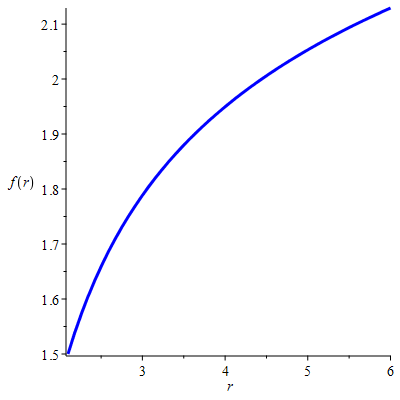}
	\caption{The conformal factor $f(r)$ against $r$ for the elastic RN solution considering $p_{t}=0$, $Q=1$, $M=1$ and ICs: $f(2.1)=1.5$, $f'(2.1)=0.5$}
	\label{f:3.6}
\end{figure}

In this case, Eq. (\ref{e:3.58}) reduces to
\begin{align}
\nonumber &-6r^{2}ff''Q^{2}+3r^{2}f'^{2}Q^{2}+6rff'Q^{2}-3f^{2}Q^{2}+12r^{3}ff''M-6r^{3}f'^{2}M-6r^{4}ff''+3r^{4}f'^{2}-6r^{3}ff' \\
\label{e:3.59} &\,\,\,\,= -3Q^{2}.  
\end{align}
Fig.\,\ref{f:3.6} explains the conformal factor of elastic RN solution considering zero tangential pressure. Hence it describes the numerical solution of the differential equation (\ref{e:3.59}). In the plot, we take $M=1$ and $Q=1$.
	
\subsubsection{Case 3: $\Lambda\neq0$ and $Q=0$}\label{sss:3.23}
\begin{figure}
	\centering
	\includegraphics[width=3.8 in]{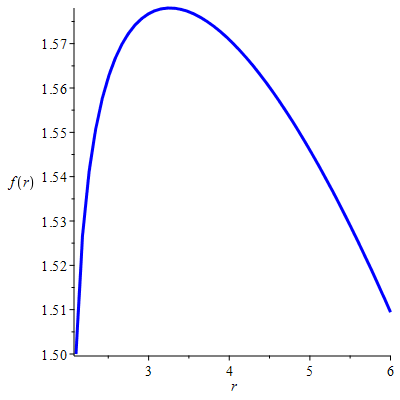}
	\caption{The conformal factor $f(r)$ against $r$ for the elastic SdS solution considering $p_{t}=0$, $\Lambda=0.01$, $M=1$ and ICs: $f(2.1)=1.5$, $f'(2.1)=0.5$}
	\label{f:3.7}
\end{figure}

Now, Eq. (\ref{e:3.58}) reduces to
\begin{equation}
\label{e:3.60} 2r^{6}ff''\Lambda-r^{6}f'^{2}\Lambda+6r^{5}ff'\Lambda+3r^{4}f^{2}\Lambda+12r^{3}ff''M-6r^{3}f'^{2}M-6r^{4}ff''+3r^{4}f'^{2}-6r^{3}ff' = 3\Lambda r^{4}f^{4}. 
\end{equation}
Fig.\,\ref{f:3.7} demonstrates conformal factor of the elastic SdS solution considering zero tangential pressure. And hence describes the numerical solution of Eq. (\ref{e:3.60}). Here we consider $\Lambda=0.01$ and $M=1$.
	
\subsubsection{Case 4: $\Lambda=0$ and $Q=0$}\label{sss:3.24}
\begin{figure}
	\centering
	\includegraphics[width=3.8 in]{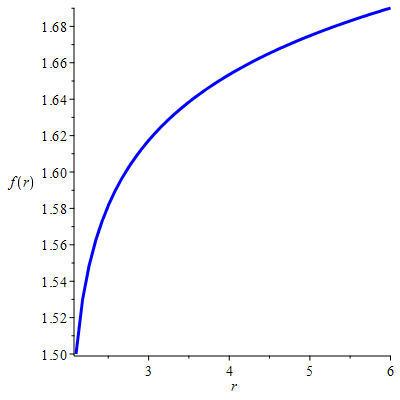}
	\caption{The conformal factor $f(r)$ against $r$ for the elastic Schwarzschild solution considering $p_{t}=0$, $M=1$ and ICs: $f(2.1)=1.5$, $f'(2.1)=0.5$}
	\label{f:3.8}
\end{figure}

Eq. (\ref{e:3.58}) reduces to
\begin{equation}\label{e:3.61}
12r^{3}ff''M-6r^{3}f'^{2}M-6r^{4}ff''+3r^{4}f'^{2}-6r^{3}ff' = 0.
\end{equation}
Fig.\,\ref{f:3.8} explains the numerical solution of Eq. (\ref{e:3.61}). It demonstrates the conformal factor of elastic Schwarzschild solution considering zero tangential pressure. It is plotted by taking $M=1$.
\par From Figs.\,\ref{f:3.1}-\ref{f:3.8}, we notice that eigenvalues $\alpha$ and $\beta$ of the metric $k^{l}_{i}$ are positive in the chosen range of $r$. Further, it is clear that the obtained numerical solutions in all the above cases describe the conformal factor and hence eigenvalues $\alpha$ and $\beta$ (see Eq. (\ref{e:3.17})).

\par The particular form of Eq. (\ref{e:3.41}) is considered in the following examples:

\subsection{Example 3}\label{ss:3.3}
Consider an elastic fluid with $p_{r}\neq0$ and $p_{t}\neq0$. Further, we take $C=2$. Then one obtains $$f^{2}(r)=\bar{a}^{2}\,\frac{(3r^{2}-6rM-\Lambda r^{4}+3Q^{2})^{2}}{r^{2}}\,(\tilde{r})^{-2}\,(\tilde{r}')^{4}.$$ Now, we consider the function $\tilde{r}(r)$ such that $\frac{(\tilde{r}')^{4}}{\tilde{r}^{2}}=C_{1}$, where $C_{1}\in\mathbb{R}^{+}$. The solution of the ordinary differential equation $\frac{(\tilde{r}')^{4}}{\tilde{r}^{2}}=C_{1}$ is given by
\begin{equation}\label{e:3.42}
\tilde{r}(r) = \frac{\sqrt{C_{1}}}{4}(C_{2}-r)^{2},
\end{equation}
where $C_{2}\in\mathbb{R}$ (see Fig.\,\ref{f:3.13}). Now, Eq. (\ref{e:3.41}) becomes
\begin{equation}\label{e:3.43}
f^{2}(r) = b\,\frac{(3r^{2}-6rM-\Lambda r^{4}+3Q^{2})^{2}}{r^{2}},
\end{equation}
where $b\in\mathbb{R}^{+}$. In this case, the following line-element describes the metric of elastic fluid:
\begin{equation}
\label{e:3.44} ds^{2} = b\,\frac{(3r^{2}-6rM-\Lambda r^{4}+3Q^{2})^{2}}{r^{2}}\left[-\xi(r)dt^{2}+\xi(r)^{-1}dr^{2}+r^{2}d\Omega^{2}\right].
\end{equation}
\par The energy density, radial pressure, and tangential pressure are as follows, respectively:
\begin{align}
\label{e:3.45} \rho &= -\frac{8r^{2}\Lambda-3}{8\pi b(r^{4}\Lambda-3Q^{2}+6rM-3r^{2})^{2}}+\frac{\Lambda}{8\pi}+\frac{Q^{2}}{8\pi b^{2}(r^{4}\Lambda-3Q^{2}+6rM-3r^{2})^{4}}, \\
\nonumber p_{r} &= \frac{16r^{7}\Lambda^{2}+16r^{3}Q^{2}\Lambda+24r^{4}M\Lambda-39r^{5}\Lambda+24MQ^{2}-27rQ^{2}-18r^{2}M+21r^{3}}{8\pi br(r^{4}\Lambda-3Q^{2}+6rM-3r^{2})^{3}}\\
\label{e:3.46} &\,\,-\frac{\Lambda}{8\pi}-\frac{Q^{2}}{8\pi b^{2}(r^{4}\Lambda-3Q^{2}+6rM-3r^{2})^{4}}, \\
\nonumber p_{t} &= \frac{8r^{8}\Lambda^{2}-14r^{4}Q^{2}\Lambda+30r^{5}M\Lambda-21r^{6}\Lambda+18Q^{4}-30rMQ^{2}+9r^{2}Q^{2}+3r^{4}}{8\pi br^{2}(r^{4}\Lambda-3Q^{2}+6rM-3r^{2})^{3}} \hspace{0.5cm} \\
\label{e:3.47} &\,\,-\frac{\Lambda}{8\pi}+\frac{Q^{2}}{8\pi b^{2}(r^{4}\Lambda-3Q^{2}+6rM-3r^{2})^{4}}. 
\end{align}
\par
\begin{figure}
	\centering
	\includegraphics[width=3.8 in]{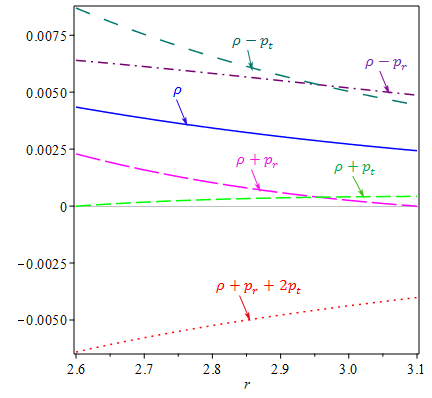}
	\caption{The energy conditions against $r$ considering $b=0.1$, $\Lambda=0.01$, $Q=2$ and $M=1$}
	\label{f:3.9}
\end{figure}

 The WEC, NEC, DEC, and SEC are plotted in Fig.\,\ref{f:3.9}, wherein we take $b=0.1$, $M=1$, $Q=2$, and $\Lambda=0.01$. The range of $r$ is chosen such that eigenvalues $\alpha$ and $\beta$ satisfy $\alpha, \beta>0$.
\par From Fig.\,\ref{f:3.9}, energy density is found to be positive. Further, it is evident from Fig.\,\ref{f:3.9} that NEC terms $\rho+p_{r}$ and $\rho+p_{t}$ are positive. And DEC terms $\rho-p_{r}$ and $\rho-p_{t}$ are also positive. Therefore, NEC, WEC, and DEC are validated. The SEC term $\rho+p_{r}+2p_{t}$ is observed to be negative. Hence SEC is violated.

\subsection{Example 4}\label{ss:3.4}
\begin{figure}
	\centering
	\includegraphics[width=3.8 in]{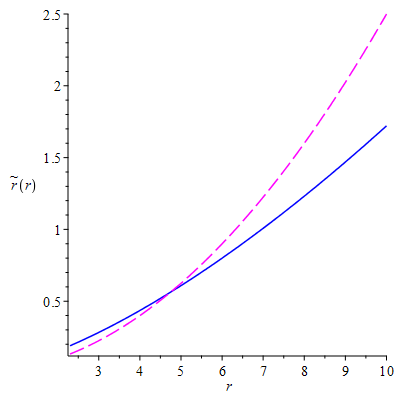}\\
	\caption{$\tilde{r}=\tilde{r}(r)$ of Example 3 for $C_{1}=0.01$ and $C_{2}=0$ (dashed line) and $\tilde{r}=\tilde{r}(r)$ of Example 4 for $C_{3}=0.01$ and $C_{4}=0$ (solid line)}
	\label{f:3.13}
\end{figure}

Here, we discuss an elastic fluid with $p_{r}\neq0$, $p_{t}\neq0$, and we take $C=3$. Then the conformal factor $f^{2}(r)$ becomes $$f^{2}(r)=\bar{a}^{2}\,\frac{(3r^{2}-6rM-\Lambda r^{4}+3Q^{2})^{\frac{3}{2}}}{r^{2}}\,(\tilde{r})^{-1}\,(\tilde{r}')^{3}.$$ We can view the function $\tilde{r}(r)$ such that $\frac{\tilde{r}'^{3}}{\tilde{r}}=C_{3}$, where $C_{3}\in\mathbb{R}^{+}$. The solution of the ordinary differential equation $\frac{\tilde{r}'^{3}}{\tilde{r}}=C_{3}$ is given by
\begin{equation}\label{e:3.48}
\tilde{r}(r) = \sqrt{\frac{8C_{3}}{27}}(r+C_{4})^{\frac{3}{2}},
\end{equation}
where $C_{4}\in\mathbb{R}$ (see Fig.\,\ref{f:3.13}). Now, Eq. (\ref{e:3.41}) becomes
\begin{equation}\label{e:3.49}
f^{2}(r) = d\,\frac{(3r^{2}-6rM-\Lambda r^{4}+3Q^{2})^{\frac{3}{2}}}{r^{2}},
\end{equation}
where $d\in\mathbb{R}^{+}$. For this case, the following line-element explains the metric of elastic fluid:
\begin{equation}
\label{e:3.50} ds^{2} = d\,\frac{(3r^{2}-6rM-\Lambda r^{4}+3Q^{2})^{\frac{3}{2}}}{r^{2}}\left[-\xi(r)dt^{2}+\xi(r)^{-1}dr^{2}+r^{2}d\Omega^{2}\right].
\end{equation}
\par $r$ is plotted versus $\tilde{r}(r)$ in Fig.\,\ref{f:3.13}. $\tilde{r}(r)$ of Example 3 is plotted for $C_{1}=0.01$ and $C_{2}=0$ (dashed line). $\tilde{r}(r)$ of Example 4 is plotted for $C_{3}=0.01$ and $C_{4}=0$ (solid line).
\par The energy density, radial pressure, and tangential pressure are as follows, respectively:
\begin{align}
\nonumber \rho &= \frac{20r^{6}\Lambda^{2}-72r^{2}Q^{2}\Lambda+132r^{3}M\Lambda-68r^{4}\Lambda+24Q^{2}-9M^{2}-30rM+15r^{2}}{32\pi d(-r^{4}\Lambda+3Q^{2}-6rM+3r^{2})^{\frac{5}{2}}}+\frac{\Lambda}{8\pi} \\
\label{e:3.51} &\,\,+\frac{Q^{2}}{8\pi d^{2}(3r^{2}-6rM-r^{4}\Lambda+3Q^{2})^{3}}, \\
\nonumber p_{r} &= -\frac{36r^{7}\Lambda^{2}+48r^{3}Q^{2}\Lambda+36r^{4}M\Lambda-80r^{5}\Lambda+72MQ^{2}-84rQ^{2}-27rM^{2}+6r^{2}M+33r^{3}}{32\pi dr(-r^{4}\Lambda+3Q^{2}-6rM+3r^{2})^{\frac{5}{2}}} \\
\label{e:3.52} &\,\,-\frac{\Lambda}{8\pi}-\frac{Q^{2}}{8\pi d^{2}(3r^{2}-6rM-r^{4}\Lambda+3Q^{2})^{3}}, \\
\nonumber p_{t} &= -\frac{20r^{8}\Lambda^{2}-48r^{4}Q^{2}\Lambda+84r^{5}M\Lambda-52r^{6}\Lambda+72Q^{4}-144rMQ^{2}+48r^{2}Q^{2}+27r^{2}M^{2}}{32\pi dr^{2}(-r^{4}\Lambda+3Q^{2}-6rM+3r^{2})^{\frac{5}{2}}} \\
\label{e:3.53} &\,\,+\frac{6r^{3}M-3r^{4}}{32\pi dr^{2}(-r^{4}\Lambda+3Q^{2}-6rM+3r^{2})^{\frac{5}{2}}}-\frac{\Lambda}{8\pi}+\frac{Q^{2}}{8\pi d^{2}(3r^{2}-6rM-r^{4}\Lambda+3Q^{2})^{3}}.
\end{align}
\par
\begin{figure}
	\centering
	\includegraphics[width=3.8 in]{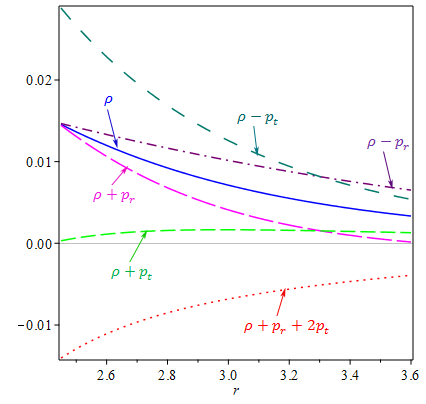}
	\caption{The energy conditions against $r$ considering $d=0.1$, $\Lambda=0.01$, $Q=2$ and $M=1$}
	\label{f:3.10}
\end{figure}

 The viability of energy conditions like WEC, NEC, DEC, and SEC is shown in Fig.\,\ref{f:3.10}. Here we consider $d=0.1$, $M=1$, $Q=2$, and $\Lambda=0.01$. The range of $r$ is chosen such that $\alpha, \beta>0$.

\par As in the previous example, from Fig.\,\ref{f:3.10}, we have found that WEC, NEC, and DEC are validated. And the SEC term $\rho+p_{r}+2p_{t}$ is observed to be negative, which implies that SEC is violated. Therefore, we deduce that NEC, WEC, and DEC are satisfied for elastic fluid with non-zero radial and tangential pressures.

\section{Matching conditions}\label{s:4}
\par $\ddot{\text{O}}$zdemir \cite{Ozdemir} considered the spherical boundary surface which divides the space-time into two distinct four-dimensional manifolds. While studying the matching problem with quadrupole moment, one can identify the location of the matching surface $\Sigma$ anywhere outside the region of the repulsive gravitational potential. Whereas in other cases, the identification of the location of $\Sigma$ is not clear \cite{Quevedo}. The matching of constant inclination surface and constant radius surface on the hypersurface $\Sigma$ is discussed in \cite{Shimano}. In \cite{Ozdemir}, one can see the matching of RNdS solution with the space-time, whose metric is as follows:
\begin{equation}\label{e:3.62}
ds^{2} = -a^{2}dt^{2}+b^{2}dR^{2}+b^{2}R^{2}d\Omega^{2},
\end{equation}
where $a=a(R, t)$ and $b=b(R, t)$.
\par To validate our constructed solution, we should match it with an exterior space-time on the spherical boundary surface $\Sigma$. In the present work, we have shown that the elastic RNdS solution (\ref{e:3.11}) with conformal factor specified in (\ref{e:3.41}) can be matched to the solution given by the metric (\ref{e:3.62}) provided we have considered the static configuration, that is $a=a(R)$ and $b=b(R)$.
\par Let $ds_{+}^{2}$ be the exterior space-time line-element expressed as
\begin{equation}\label{e:3.63}
ds_{+}^{2} = -a^{2}dt^{2}+b^{2}dR^{2}+b^{2}R^{2}d\Omega^{2}.
\end{equation}
Let $ds_{-}^{2}$ be the interior space-time line-element expressed as
\begin{equation}\label{e:3.64}
ds_{-}^{2} = f^{2}(r)\left[-\xi(r)dt^{2}+\xi(r)^{-1}dr^{2}+r^{2}d\Omega^{2}\right].
\end{equation}
\par The spherical boundary surface $\Sigma$ is considered as the boundaries of exterior and interior space-times at $\mathcal{R}_{\Sigma}=r_{\Sigma}=bR_{\Sigma}=\text{constant}$. Let $ds_{\Sigma}^{2}$ be the line-element of the spherical boundary surface $\Sigma$ of constant radius $\mathcal{R}_{\Sigma}$. The symbols $+$ and $-$ indicate exterior and interior quantities, respectively. Let $n^{i-}$ and $n^{i+}$ denote the normal vectors to the spherically symmetric boundary surface $\Sigma$ and are given by
\begin{equation}\label{e:3.65}
n^{i+} = \frac{1}{b}\,\partial_{R},\,\,\,\,n^{i-}\,=\,\frac{\sqrt{\xi(r)}}{f}\,\partial_{r}.
\end{equation}
Let $T\Sigma^{\pm}$ be the tangent space to the spherically symmetric boundary surface $\Sigma$ and is as follows:
\begin{equation}\label{e:3.66}
T\Sigma^{\pm} =\,<e_{1}^{\pm},\,e_{2}^{\pm},\,e_{3}^{\pm}>,
\end{equation}
where $e_{1}^{\pm}\,=\,\partial_{t},\,e_{2}^{\pm}\,=\,\partial_{\theta},\,e_{3}^{\pm}\,=\,\partial_{\phi}$. Let $q_{\mu\nu}^{+}, q_{\mu\nu}^{-}$ be the first fundamental forms at $\Sigma$. The expression of first fundamental forms at $\Sigma$ is given by
\begin{equation}\label{e:3.67}
q_{\mu\nu}^{\pm} = e_{\mu}^{\pm i}e_{\nu}^{\pm j}g_{ij}^{\pm},
\end{equation}
$\mu, \nu=1, 2, 3$. The first fundamental forms for both interior space-time and exterior space-time at the spherical boundary surface $\Sigma$ are
\begin{align}
\label{e:3.68} d\sigma_{-}^{2} &= -f^{2}(r)\xi(r)dt^{2}+r^{2}f^{2}(r)d\Omega^{2}, \\
\label{e:3.69} d\sigma_{+}^{2} &= -a^{2}(R)dt^{2}+R^{2}b^{2}(R)d\Omega^{2}.
\end{align}
In other words, $d\sigma_{+}^{2}=ds_{+}^{2}|_{\Sigma}$ and $d\sigma_{-}^{2}=ds_{-}^{2}|_{\Sigma}$, that is, all the quantities should be calculated on the boundary surface $\Sigma$. The equality of the first fundamental forms, $q_{\mu\nu}^{+}=q_{\mu\nu}^{-}$, is explained by the first matching conditions. In other words, $ds_{\Sigma}^{2}=ds_{+}^{2}|_{\Sigma}=ds_{-}^{2}|_{\Sigma}$. Therefore, we have
\begin{align}
\label{e:3.70} a^{2}(R) &\overset{\Sigma}{=} f^{2}(r)\xi(r), \\
\label{e:3.71} R^{2}b^{2}(R) &\overset{\Sigma}{=} r^{2}f^{2}(r).
\end{align}
Here, equality holds true only on the spherical boundary surface $\Sigma$. This is shown by the symbol $\overset{\Sigma}{=}$ in the expressions (\ref{e:3.70})-(\ref{e:3.71}). The continuity of the space-time metric across the spherical boundary surface $\Sigma$ is confirmed by the first matching conditions \cite{Mena}.
\par Let $H_{\mu\nu}^{+}, H_{\mu\nu}^{-}$ be the extrinsic curvatures associated with $\Sigma$. The expression of second fundamental forms at $\Sigma$ is given by
\begin{equation}\label{e:3.72}
H_{\mu\nu}^{\pm} = -n_{i}^{\pm}e_{\mu}^{\pm j}\nabla_{j}^{\pm}e_{\nu}^{\pm i}.
\end{equation}
The second matching conditions give the continuity of the extrinsic curvature $H_{\mu\nu}$ at the spherical boundary surface $\Sigma$. From \cite{Mars}, we have found that matching conditions allow two arbitrary discontinuities in the Weyl tensor and four in the matter tensor at null points of the hypersurface. Further, we have observed that at non-null points, matching conditions allow six independent discontinuities in the Riemann tensor, and these are the matter discontinuities. Let $\zeta^{\mu}=(t, \theta, \phi)$ be the coordinates on $\Sigma$, and $x^{\pm l}$ are the coordinates of both exterior and interior space-times. The extrinsic curvature of $\Sigma$ is also given by \cite{Eisenhart}
\begin{equation}\label{e:3.73}
H_{\mu\nu}^{\pm} = -n_{l}^{\pm}\left(\frac{\partial^{2}x^{l}_{\pm}}{\partial\zeta^{\mu}\partial\zeta^{\nu}}+\Gamma^{l}_{pq}\frac{\partial x^{p}_{\pm}}{\partial\zeta^{\mu}}\frac{\partial x^{q}_{\pm}}{\partial\zeta^{\nu}}\right),
\end{equation}
where $l, p, q=0, 1, 2, 3$. The second fundamental forms have the following non-zero components for the interior and exterior space-times at $\Sigma$:
\begin{align}
\label{e:3.74} H_{tt}^{-} &= \left\{\frac{r(rf'+f)\Lambda}{3}+\frac{(f-rf')Q^{2}}{r^{3}}+\frac{(2rf'-f)M}{r^{2}}-f'\right\}\sqrt{\xi(r)}, \\
\label{e:3.75} H_{\theta\theta}^{-} &= r(rf'+f)\sqrt{\xi(r)}, \\
\label{e:3.76} H_{tt}^{+} &= -\frac{aa'}{b}, \\
\label{e:3.77} H_{\theta\theta}^{+} &= R(Rb'+b).
\end{align}
The equality, $H_{\mu\nu}^{+}=H_{\mu\nu}^{-}$, leads to the following relations:
\begin{align}
\label{e:3.78} \frac{aa'}{b} &\overset{\Sigma}{=} \left\{f'-\frac{(2rf'-f)M}{r^{2}}-\frac{r(rf'+f)\Lambda}{3}-\frac{(f-rf')Q^{2}}{r^{3}}\right\}\sqrt{\xi(r)}, \\
\label{e:3.79} R(Rb'+b) &\overset{\Sigma}{=} r(rf'+f)\sqrt{\xi(r)}.
\end{align}
The radial pressure for the exterior space-time (\ref{e:3.63}) can be expressed as
\begin{equation}\label{e:3.64(c)}
p_{r}^{+} = -\frac{Rab'^{2}+2Rba'b'+2abb'+2b^{2}a'}{8\pi Rab^{4}}-\frac{\Lambda}{8\pi}-\frac{Q^{2}}{8\pi R^{4}b^{4}}.
\end{equation}
\par Using Eqs. (\ref{e:3.70}), (\ref{e:3.71}), (\ref{e:3.78}) and (\ref{e:3.79}), one can evaluate the radial pressure, $p_{r}^{-}$ at $\Sigma$, and the resultant expression is given by
\begin{align}
\nonumber p_{r}^{-} &\overset{\Sigma}{=} \frac{r^{6}f'^{2}\Lambda+2r^{5}ff'\Lambda+r^{4}f^{2}\Lambda-3r^{2}f'^{2}Q^{2}-2rff'Q^{2}+f^{2}Q^{2}+6r^{3}f'^{2}M+6r^{2}ff'M-3r^{4}f'^{2}-4r^{3}ff'}{8\pi r^{4}f^{4}}  \\
\label{e:3.81}  &\,\,-\frac{Q^{2}}{8\pi r^{4}f^{4}}-\frac{\Lambda}{8\pi}.
\end{align}
From Eqs. (\ref{e:3.31}), (\ref{e:3.64(c)}) and (\ref{e:3.81}), we conclude that
\begin{equation}\label{e:3.82}
p_{r}^{+} \overset{\Sigma}{=} p_{r}^{-}.
\end{equation}
The mass $m^{+}$ of the exterior solution can be defined as shown below:
\begin{equation}\label{e:3.82(a)}
m^{+} = 4\pi\int\limits^{R}_{0}\rho^{-}\bar{R}^{2}b^{2}(\bar{R}b)'d\bar{R}.
\end{equation}
By making use of Eq. (\ref{e:3.82(a)}), the mass $m^{+}$ is obtained to be
\begin{equation}\label{e:3.83}
m^{+} = \frac{1}{2}\left(\frac{1}{3}\Lambda R^{3}b^{3}-\frac{Q^{2}}{Rb}+\frac{R^{3}b'^{2}}{b}+2R^{2}b'\right).
\end{equation}
The mass $m^{-}$ of the interior solution can be evaluated by applying the definition given below:
\begin{equation}\label{e:3.83(a)}
m^{-} = 4\pi\int\limits^{r}_{0}\rho\bar{r}^{2}f^{2}(\bar{r}f)'d\bar{r},
\end{equation}
where $\rho$ is given by Eq. (\ref{e:3.30}). We obtain
\begin{equation}\label{e:3.84}
m^{-} = \frac{\Lambda r^{3}f^{3}}{6}-\frac{rf}{2}-\frac{Q^{2}}{2rf}+\frac{(3r^{2}-6rM-\Lambda r^{4}+3Q^{2})(rf'+f)^{2}}{6rf}.
\end{equation}
The formulae (\ref{e:3.82(a)}) and (\ref{e:3.83(a)}) are compatible with the definitions given in \cite{Brito4,Brito,Wald2,Florides} for calculating the mass. Under the Eqs. (\ref{e:3.71}) and (\ref{e:3.79}), the expression for $m^{-}$ at $\Sigma$ gives
\begin{equation}\label{e:3.85}
m^{-} \overset{\Sigma}{=} m^{+}.
\end{equation}

\section{Conclusions and future perspectives}\label{s:5}
\par In this paper, we have constructed the elastic solution of the EFEs by applying conformal transformations to the RNdS solution. The problem has been discussed with the static configuration. Thus, the conformal factor depends on the radial coordinate $r$ only. In other words, the conformal factor is independent of the time coordinate $t$. As a consequence, by solving (\ref{e:3.39}), we have obtained the expression for the conformal factor (see (\ref{e:3.40})). Whereas in the case of non-static configuration, the conformal factor depends on both $r$ and $t$. So we will obtain PDE. Due to its complexity, only a specific family of solutions can be determined for the conformal factor. Hence, the present work deals with the static configuration only. We will construct a non-static conformal elastic matter solution of the EFEs from the RNdS space-time metric in the future. Further, discussing the matching problem for the non-static configuration with the exterior space-time and investigating the physical viability of the subfamily of static solutions of this specific family of non-static solutions are the possible directions for further research. Further, we will carry out a study on trapped surfaces in the static conformal elastic solutions also in the future.

\par Four examples have been analyzed: in first-one, we have discussed the elastic fluid having non-zero tangential pressure and zero radial pressure; the elastic fluid with non-zero radial pressure and zero tangential pressure is studied in the second example; and in both the third and fourth examples, we have examined the elastic fluid with non-zero radial and tangential pressures. Further, the viability of energy conditions is analyzed for both the third and fourth examples. And found that WEC, NEC, and DEC are satisfied in both. We have investigated the matching problem with the exterior space-time across a spherical boundary surface and shown that the elastic RNdS solution can be matched to the exterior solution.

\par For constructed elastic solutions, some difficulties arise in investigating the stability issue due to the coupling between the electromagnetic and the gravitational perturbations of the solution. In \cite{Giorgi}, the author tried to resolve the stability issue due to the coupling between the electromagnetic and the gravitational perturbations of RN solution considering a small electric charge, that is, the case of $|Q|\ll M$. Therefore, we leave the stability analysis of elastic solutions of EFEs as a perspective.

\par By calculating the binding energies and the energies accumulated in the gravitational field, one can analyze the stability of elastic solutions of EFEs concerning their binding energies. We hope to discuss it in the future. Further, stability analysis through binding energy would be exciting since it will help in discussing neutron stars and white dwarfs. Using the technique of \cite{Nashed} also, we have planned to analyze the stability of constructed elastic solutions of EFEs in future work.

\par One can conclude that the avenue adopted in the present work and Refs. \cite{Brito4,Brito3,Brito2,Brito}, allows constructing the new exact solutions of the EFEs with the elastic matter.
	
\section*{Acknowledgments}
The author H. M. Manjunatha is very much grateful to Karnataka Science and Technology Promotion Society (KSTePS), Department of Science and Technology (DST), Govt. of Karnataka (Award Letter No. OTH-04:\,2018-19), for awarding DST-Ph.D. Fellowship.

\bibliographystyle{unsrt}

\end{document}